\documentclass[prl,amsmath,aps,twocolumn,superscriptaddress]{revtex4}
\usepackage{amsfonts}
\usepackage{graphicx}
\usepackage{times}
\newcommand{\tr}{{\rm Tr}}
\begin{document}
%\title{Commutativity-preserving quantum channels and quantum correlations}
\title{Quantum channels that preserve the commutativity}
\author{Sixia Yu}
\affiliation{Centre for Quantum Technologies and Physics Department,
National University of Singapore, 2 Science Drive 3, Singapore
117542}
\affiliation{Hefei National Laboratory for Physical Sciences at
Microscale and Department of Modern Physics, University of Science
and Technology of China, Hefei, Anhui 230026, P.R. China}
\author{Chengjie Zhang}
\affiliation{Centre for Quantum Technologies and Physics Department,
National University of Singapore, 2 Science Drive 3, Singapore
117542}
\author{Qing Chen}
\affiliation{Centre for Quantum Technologies and Physics Department,
National University of Singapore, 2 Science Drive 3, Singapore
117542}
\author{C.H. Oh}
\affiliation{Centre for Quantum Technologies and Physics Department,
National University of Singapore, 2 Science Drive 3, Singapore
117542}

\begin{abstract}
%The quantum  correlations, including entanglement and quantum discord, are useful resources.
We identify and characterize all the local quantum channels that preserves the set of classical states, i.e., does not create any quantum correlations. At first we show that the quantum correlations cannot be created from without if and only if the local quantum channel preserves the commutativity, i.e., the images of any two commuting states also commute. And then we provide an operational necessary and sufficient criterion for a known quantum channel to preserve the commutativity as well as a single observable to witness an arbitrary unknown commutativity-preserving channel. All the distance-based measures for quantum correlations, e.g., the geometric measure, are non-increasing while the quantum discord defined by von Neumann measurements can be increasing or decreasing under local commutativity-preserving channels.
\end{abstract}

\maketitle

%\paragraph{Introduction---}

Quantum correlations, including the entanglement and the quantum discord, are useful resources that play a fundamental role in various quantum informational processes. Quantum correlations beyond entanglement, i.e., quantum correlations found in separable states, e.g., as quantified by quantum discord \cite{discord1,discord2}, are argued to be responsible for the speedup in certain quantum computational tasks such as deterministic quantum computation with one qubit \cite{dqc1,datta}. The operational interpretations via state merging \cite{int1} for the quantum discord establish firmly the status of quantum discord as a useful resource beside the entanglement. Also the quantum discord is related to the completely positive maps and quantum phase transitions in certain physical systems. As a resource it is therefore important to understand how the quantum correlations behave under local noises or operations.

As is well-known, the entanglement is non-increasing under local operations and classical communications (LOCC). Especially the entanglement cannot be created from a separable state using only LOCCs. This property of entanglement is characteristic for its various quantitative measures. It is natural to ask what kinds of local operations that play the role of LOCCs for quantum correlations and especially what kind of local operations that do not create the quantum correlations as quantified by quantum discord.
Operations allowed by quantum mechanics are quantum channels, i.e, all possible dynamical processes described by trace-preserving completely positive (CP) maps. In the case of qubit channels this problem has been solved in \cite{2cp}: qubit channels that do not create the quantum discord are either unital, i.e., mapping identity operator to identity operator, or semi-classical, i.e, nullifying quantum correlations in any state. For higher dimensional systems there are examples of unital channels that are able to create quantum correlations. The problem of characterizing all the quantum channels for higher dimensional system that do not create any quantum correlations is left open.

Here we shall resolve this problem in its full generosity. At first we show that a quantum channel does not create quantum correlation if and only if the given channel preserves the commutativity. And then we present an operational sufficient and necessary criterion for a known channel to be commutativity-preserving (CoP). Also a single observable is proposed to witness the commutativity-preserving property of an arbitrarily unknown quantum channel. As an example, we show that the mixing of a given CoP channel with the identical channel  preserves still the commutativity if and only if the given channel is a cloning channel. Finally  the behaviors of various quantitative measures of quantum correlations, especially the quantum discord, under CoP channels are discussed.

Although there are many different measures to quantify the quantum correlations beyond entanglement, bipartite classical states as well as quantum-classical or classical-quantum states are unambiguously defined. In what follows we shall use the languages of quantum discord. A bipartite state $\varrho_{AB}$ is said to be a classical-quantum state if and only if it has a vanishing $A$-discord, i.e., there exist an orthonormal basis $\{|k\rangle\}$ for subsystem $A$, a probability distribution $p_k$, and a set of density matrices of $\sigma_k^B$ of subsystem $B$ such that
\begin{equation}\label{zero}
\varrho_{AB}=\sum_kp_k|k\rangle\langle k|_A\otimes \sigma_k^B.
\end{equation}
The classical-quantum state can be defined similarly, i.e.,  with a vanishing $B$-discord, and a classical state is both classical-quantum and quantum-classical, i.e. states with vanishing $A$-discord and $B$-discord. Operationally, a given bipartite state has zero $A$-discord if and only if $\tr_B(\varrho_{AB}\lambda_\mu^B)$ are mutually commuting where $\{\lambda_\mu^B\}$ is an arbitrary operator basis for subsystem $B$ \cite{oc}. For an unknown state with four copies a single observable can witness the nonzero quantum discord \cite{cj,yu1}. Our first result concerns what kinds of local quantum operations preserve the set of states with zero $A$-discord.

{\it Theorem 1 } For a bipartite system $AB$ a local quantum channel $\Lambda$ acting on subsystem $A$ preserves the set of states with vanishing $A$-discord, i.e., the image of any zero $A$-discord state still has a zero $A$-discord, if and only if the channel $\Lambda$  preserves the commutativity, i.e., for any two density matrices $\varrho,\sigma$ it holds
\begin{equation}
[\varrho,\sigma]=0 \Rightarrow [\Lambda(\varrho),\Lambda(\sigma)]=0.
\end{equation}

{\it Proof } Consider a bipartite state $\varrho_{AB}$ with a zero A-discord as given in Eq.(\ref{zero}) and a local quantum channel $\Lambda$ acting only on subsystem $A$ that preserves the commutativity. Since all projections $\{|k\rangle\langle k|\}$ in an arbitrary basis  commute their images $\Lambda(|k\rangle\langle k|)$ are also mutually commuting. As a result the image of the bipartite $\varrho_{AB}$ has also a
zero $A$-discord according to the operational criterion \cite{oc}. If the local quantum channel $\Lambda$ does not preserve the commutativity then there exist two commuting states $\varrho,\sigma$ of subsystem $A$ such that $[\Lambda(\varrho),\Lambda(\sigma)]\ne 0$.
While the state $\varrho_{AB}=\varrho_A\otimes |0\rangle\langle0|+\sigma_A\otimes|1\rangle\langle1|$ has a zero $A$-discord, its image $\Lambda\otimes\mathcal I(\varrho_{AB})$ has a nonzero $A$-discord, also thanks to the operational criterion \cite{oc}. That is to say the quantum channel $\Lambda$ that does not preserve the commutativity is able to create nonzero quantum discord from a state with a zero quantum discord.\hfill Q.E.D.

In the following we shall characterize all the CoP channels for qudit, i.e., quantum system with $d$ energy levels.
Let  $\{|k\rangle\}_{k=1}^d$ be the computational basis and $\{\lambda_\mu|0\le\mu \le D:=d^2-1\}$ be a set of orthonormal Hermitian operator basis for qudit with $\lambda_0$ being identity and $\tr(\lambda_\mu\lambda_\nu)=d\delta_{\mu\nu}$. The structure constants
$
f_{\mu\nu \tau}=-i\tr\big([\lambda_\mu,\lambda_\nu]\lambda_\tau\big)/{d^2}$ with $\mu,\nu,\tau=0,1,\ldots,D$
are totally antisymmetric with respect to three indices.
A quantum channel $\Lambda$, i.e., a trace-preserving completely positive map, can be represented in three equivalent ways: the operator-sum representation, i.e., Kraus operators, the unitary representation on system plus environment, and a two-qudit state $R_\Lambda=\Lambda\otimes \mathcal I(\Phi)$ where $\Phi=|\Phi\rangle\langle\Phi|$ is the projection of the maximal entangled state $|\Phi\rangle=\sum_{n=1}^{d}|nn\rangle_{AB}$ (not normalized) in which qudit $A$ is called as the system qudit while the qudit $B$ is called as the reference qudit.

Consider four copies of the two-qudit state $R_\Lambda=\Lambda\otimes \mathcal I(\Phi)$ corresponding to a given channel $\Lambda$ and label four system qudits that the channels act on by $A_i$ and four reference qudits by $B_i$ with $i=1,2,3,4$. Let $V_{ij}$ be the swapping operator acting on the system (or reference) qudits $i$ and $j$ and denote by $X=V_{12}V_{23}V_{34}$ the cyclic permutation of four  qudits. The quantum discord witness introduced in \cite{yu1} is an collective observable on four copies
\begin{equation}
W=\frac{X_A+X_A^\dagger}2\otimes (V_{12}V_{34}-V_{13}V_{24}).
\end{equation}
The 2-qudit state $R_\Lambda$ has a vanishing $A$-discord if and only if $\tr(R_\Lambda^{\otimes 4}W)=0$ \cite{yu1}. For an unknown 2-qudit state with four copies we have only to measure the quantum discord witness $W$ to see whether the state has a vanishing discord or not. Also we introduce another collective observable
\begin{equation}
Z=\frac{X_A+X_A^\dagger}2\otimes V_{14}V_{23}(V_{13}+V_{24}-V_{12}-V_{34})
\end{equation}
and denote $L=-(I_A\otimes  V_{14}V_{23})Z$.
We note that all these collective observables, especially $W$ and $Z$, can be measured by introducing some auxiliary qubits, some controlled swapping gates with qubits as sources, and qubit measurements. Now we are ready to formulate our main results:

{\it Theorem 2 } The following statements are equivalent:
\begin{itemize}
\item[i)] The qudit channel $\Lambda$ preserves the commutativity.
\item[ii)] Given four copies of $R_\Lambda=\Lambda\otimes\mathcal I(\Phi)$  it holds
\begin{equation}
d\tr\left(R^{\otimes 4}_\Lambda W\right)=\tr\left(R^{\otimes 4}_\Lambda Z\right).\end{equation}
\item[iii)] In a given orthonormal Hermitian operator basis $\{\lambda_\mu\}$ with $\lambda_0$ being identity for arbitrary $\mu,\nu=0,1,\ldots, D$ it holds
\begin{equation}\label{iii}
   [\Lambda(\lambda_\mu),\Lambda(\lambda_\nu)]=
   \frac12\sum_{\alpha,\beta,\tau=1}^Df_{\mu\nu\tau}f_{\alpha\beta\tau}
   [\Lambda(\lambda_\alpha),\Lambda(\lambda_\beta)].
\end{equation}
\end{itemize}

{\it Proof } By definition a channel $\Lambda$ preserves the commutativity if and only if the images of  two arbitrary commuting density matrices $[\sigma,\varrho]=0$ still commute, i.e., $[\Lambda(\varrho),\Lambda(\sigma)]=0$. Since commuting density matrices have common eigenstates, it is equivalent to  require $i[\Lambda(\hat k_U),\Lambda(\hat l_U)]=0$ for arbitrary $U$ and $k,l$ where  $\{|k\rangle\}$ is a given qudit basis and $\hat k_U=U|k\rangle\langle k|U^\dagger$. Since for a Hermitian matrix $H$ the condition $H=0$ is equivalent to $\tr (H^\dagger H)=0$, a qudit channel $\Lambda$ preserves the commutativity if and only if
\begin{eqnarray}\label{int}
0&=&-\int dU\sum_{k,l=1}^d\tr[\Lambda(\hat k_U),\Lambda(\hat l_U)]^2\cr
&=&\tr \left(R^{\otimes 4}_\Lambda \frac{X_A+X_A^\dagger}2\otimes \int dU U^{\otimes 4}\ O^T\ U^{\dagger\otimes 4} \right)\cr
&=&\frac{\tr \Big(R^{\otimes 4}_\Lambda\big((d+1)L+dW-Z\big)\Big)}{d(d+1)(d+2)}
\end{eqnarray}
where the integral is over the unique Haar measure of unitary matrices and $O=\sum_{k,l=1}^d\hat k\otimes(\hat k\otimes\hat l-\hat l\otimes\hat k)\otimes\hat l$. To obtain the second equality above we have used the properties $\Lambda(\varrho)=\tr_B(R_\Lambda \varrho_B^T)$ and $\tr(X\varrho_1\otimes\varrho_2\otimes\varrho_3\otimes\varrho_4)=\tr(\varrho_1\varrho_2\varrho_3\varrho_4)$. To calculate the integral $\bar O=\int dU U^{\otimes 4}O U^{\dagger\otimes 4}$ in the third line in the equation above we have used the fact that $\bar O=\sum_{\sigma\in S_4}O_\sigma V_\sigma$, since $[\bar O, U^{\otimes 4}]=0$ for arbitrary $U$,  for some real numbers $O_\sigma$,
where $V_\sigma$ is the $4$-qudit operator representing the permutation $\sigma\in S_4$. To compute the coefficients $O_\sigma$ we can either use the Collins-Sniady formula \cite{cs} or solve the linear equation $\tr(\bar OV_\sigma)=\tr (OV_\sigma)$ with 24 variables $\{O_\sigma|\sigma\in S_4\}$. Taking into account $\tr (OV_{12})=\tr(OV_{34})=\tr(OV_{12}V_{34})=d(d-1)$ and $\tr(OV_{13})=\tr(OV_{24})=\tr(OV_{13}V_{24})=-d(d-1)$ with $\tr (OV_\sigma)=0$ otherwise, we obtain
\begin{eqnarray}
\bar O=\frac{(d+1)\tilde L+d(V_{12}V_{34}-V_{13}V_{24})+V_{14}V_{23}\tilde L}{d(d+1)(d+2)},
\end{eqnarray}
where
$\tilde L=V_{12}+V_{34}-V_{13}-V_{24}$.
On the other hand it is rather straightforward to check the above equality by comparing the coefficients $\tr (OV_\sigma)$ with $\tr (\bar OV_\sigma)$ which should be identical.

In what follows the summation from $0$ to $D$ over repeated indices is always assumed. In a local orthogonal observable basis $\{\lambda_\mu\otimes\lambda_\nu^T\}$ for two qudits we have expansion $d\Phi=\lambda_\mu\otimes\lambda_\mu^T$ and as a result $dR_\Lambda=\Lambda(\lambda_\mu)\otimes\lambda_\mu^T$. Straightforward calculations yield $2d^2\tr\left(R^{\otimes 4}_\Lambda W\right)=-\tr(\Lambda_{\mu\nu}\Lambda_{\mu\nu})$ and $
-4d\tr\left(R^{\otimes 4}_\Lambda Z\right)=f_{\mu\nu\tau}f_{\alpha\beta\tau}\tr\left(\Lambda_{\mu\nu}
\Lambda_{\alpha\beta}\right),$
where $\Lambda_{\mu\nu}=[\Lambda(\lambda_\mu),\Lambda(\lambda_\nu)]$. We obtain
\begin{eqnarray}\label{dt}
\delta_\Lambda&:=&\tr\left(R^{\otimes 4}_\Lambda(dW-Z)\right)\cr &=&-\frac{1}{2d}\sum_{\mu,\nu=0}^D\tr\left(\Lambda_{\mu\nu}-\frac12 f_{\mu\nu\tau}f_{\alpha\beta\tau}\Lambda_{\alpha\beta}\right)^2
.
\end{eqnarray}
In the above calculations the identity $f_{\mu\nu\tau}f_{\mu\nu\tau^\prime}=2\delta_{\tau\tau^\prime}$ is useful.
We note that each terms in the above summation is non-positive because $\Lambda^\dagger_{\mu\nu}=-\Lambda_{\mu\nu}$ and thus $\delta_{\Lambda}\ge0$. Furthermore by neglecting all the terms $\mu\nu\ne0$ in the above equation, we obtain immediately $\delta_\Lambda\ge-\sum_{\mu}\tr\Lambda_{\mu 0}^2/d=\tr\left(R^{\otimes 4}_\Lambda L\right)$. Now  it is easy to see that Eq.(\ref{int}) holds true if and only if $\delta_\Lambda=0$, which is exactly the second statement and leads to the third statement due to Eq.(\ref{dt}). \hfill Q.E.D.

Some remarks are in order. Firstly, it seems quite natural to regard the quantity $\delta_\Lambda$ as the degree of commutativity preserving for a given channel $\Lambda$ because it has some nice properties. First, so-defined CoP degree is nonnegative and becomes zero if and only if the channel preserves the commutativity.  Second, it can be readily calculated for a given channel or measured for an unknown channel provided four uses of the channel on a maximally entangled state $\Phi$. Secondly here we have concerned mainly in bipartite states and it is straightforward to see that local CoP channels also preserve multipartite classical correlations, with or without classical communications. However we leave the problem open whether all classical correlation preserving channels can arise in this way.

Thirdly, in the case of qubit the condition Eq.(\ref{iii}) becomes identities for  $\mu,\nu\ne 0$  because of the identity $\sum_{\tau=1}^3 f_{\mu\nu\tau}f_{\alpha\beta\tau}=\delta_{\mu\alpha}\delta_{\nu\beta}-\delta_{\mu\beta}\delta_{\nu\alpha}$.
Thus the necessary and sufficient condition for CoP becomes simply
$
[\Lambda(I),\Lambda(\lambda_\nu)]=0
$,
recalling that $\lambda_0=I$ is the identity by definition. Its necessity is obviously since the identity $I$ commute with all operators. To see its sufficiency we note that the condition $\Lambda_{\mu0}=0$  leads to either $\Lambda(I)=I$, i.e., the channel is unital, or $\Lambda(I)=I+\vec m\cdot\vec \sigma$ for some nonzero Bloch vector $\vec m$. From condition $\Lambda_{\mu0}=0$ it follows that $[\Lambda(\varrho),\vec m\cdot\vec\sigma]=0$ which leads to $\Lambda(\varrho)\propto \vec m\cdot\vec\sigma$ in the case of qubit, which is not necessarily true if $d\ge 3$. That is to say the channel is a semi-classical channel that is defined by
$ \mathcal B(\varrho)=\sum_{k}\tr(M_k\varrho)|\phi_k\rangle\langle \phi_k|$ with $\{M_k\}$ being a $d$-outcome POVM and $\{|\phi_k\rangle\}$ an arbitrary basis. In fact the semi-classical channel nullifies the quantum discord of all the states, i.e., it brings any bipartite state to a classical-quantum state.  Thus in the case of qubits, all the CoP channels are either unital, i.e., $\Lambda(I)=I$ or semi-classical, which reproduces the results for qubit channels in \cite{2cp}. The cloning channel
$\mathcal C(\varrho)\propto  I\tr\varrho+c\varrho$ is obviously another example of CoP channel for all $0\le c\le 1$. One special case is the identity channel $\mathcal I(\varrho)=\varrho $.

As an example of non-CoP channel we consider the channel
$\mathcal H(\varrho)=({I\tr\varrho -i[H,\varrho]})/d$, which is referred to as Hamiltonian channel here,
with $H$ being an arbitrary nonzero Hermitian qudit operator with zero trace. The channel is positive if $\tr H^2\le 1/2$, since $-\tr [H,\varrho]^2\le 2\tr H^2$, and completely positive if $\tr H^2\le 1/d$, since $dR_{\mathcal H}$ has eigenvalues $1\pm\sqrt{d\tr H^2}$. Straightforward calculation yields
$d^4\delta_{\mathcal H}=(d^2-6)(\tr H^2)^2+d(d^2-2)\tr H^4$. Since $d\tr H^4\ge (\tr H^2)^2$, in which the equality holds true for arbitrary traceless qubit operator $H$, we always have $d^4\delta_{\mathcal H}\ge 2(d^2-4)(\tr H^2)^2>0$ in the case of $d\ge 3$. Therefore all Hamiltonian channels for $d\ge 3$ do not preserve the commutativity. And in the case of qubit $\delta_{\mathcal H}=0$ and thus the Hamiltonian channel preserves the commutativity.

By a mixing of two channels $\Lambda_1$ and $\Lambda_2$ we refer to the channel $p\Lambda_1+(1-p)\Lambda_2$ for some $1\le p\ge 0$. Firstly let us consider the mixing $\Lambda_p=p\mathcal I+\bar p\mathcal H$ of the identical channel with a Hamiltonian channel. By using the Jacobi identity $[\lambda_\mu,[H,\lambda_\nu]]-[\lambda_\nu,[H,\lambda_\mu]]=[H,[\lambda_\mu,\lambda_\nu]]$ we obtain $\delta_{\Lambda_p}=\bar p^4\delta_{\mathcal H}$. Thus as long as $H\ne 0$, $p\ne 1$, and $d\ge 3$ we have $\delta_{\Lambda_p}>0$, i.e., the mixing $\Lambda_p$ of a Hamiltonian channel with the identity channel preserves the commutativity if and only if $d=2$ or $H=0$ or $p=1$.

Now let us consider a nontrivial mixing $\Lambda^\prime=p\mathcal I+\bar p\Lambda$ of the identity channel $\mathcal I$ and an arbitrary CoP channel $\Lambda$ with $1>p>0$ and $\bar p=1-p$. By denoting $\Lambda^\prime_{\mu\nu}=[\Lambda^\prime(\lambda_\mu),\Lambda^\prime(\lambda_\nu)]$ we have $\Lambda^\prime_{\mu\nu}=p^2[\lambda_\mu,\lambda_\nu]+\bar p^2\Lambda_{\mu\nu}+p\bar p S_{\mu\nu}$
where $S_{\mu\nu}=[\lambda_\mu,\Lambda(\lambda_\nu)]-[\lambda_\nu,\Lambda(\lambda_\mu)]$. To evaluate the CoP degree of $\Lambda^\prime$, we note at first that $\delta_{\mathcal I}=\delta_\Lambda=0$ since both $\mathcal I$ and $\Lambda$ are CoP channels. In terms of real coefficients $\Lambda_\mu^\alpha=\tr(\Lambda(\lambda_\mu)\lambda_\alpha)$  the CoP degree $\delta_{\Lambda^\prime}$ turns out to be a quadratic from. Straightforward calculations yield
\begin{equation}
\delta_{\Lambda^\prime}=p^2\bar p^2\langle\Lambda|(2-\mathbf K)(1+\mathbf K)|\Lambda\rangle
\end{equation}
 where we have denoted by $|\Lambda\rangle=\sum_{\mu\alpha}\Lambda_\mu^\alpha|\mu,\alpha\rangle$ a two-quDit (qudit with $D=d^2-1$ levels) state and by $\mathbf K$ a two-quDit observable given by $\langle\mu,\alpha|\mathbf K|\beta,\nu\rangle=f_{\mu\nu\tau}f_{\alpha\beta\tau}$.
In order to find out the eigenvalue of $\mathbf K$ let us look at its square $
\mathbf K^2=4\Psi+\mathbf E+\mathbf F+4\mathbf{P}/{d^2}$
where $\Psi=|\Psi\rangle\langle \Psi|/D$ is the projector to the pure quDit state $|\Psi\rangle=\sum_\mu|\mu,\mu\rangle$, $\mathbf F=\sum_{\tau}|f_\tau\rangle\langle f_\tau|/2$, $\mathbf E=d^2\sum_{\tau}|e_\tau\rangle\langle e_\tau|/(2(d^2-4))$ with $|f_\tau\rangle=\sum_{\mu\nu}f_{\mu\nu\tau}|\mu,\nu\rangle$ and $|e_\tau\rangle=\sum_{\mu\nu}\tr(\{\lambda_\mu,\lambda_\nu\}\lambda_\tau)|\mu,\nu\rangle/d^2$ being orthogonal to each other for $\tau=1,2,\ldots,D$. Moreover $\mathbf P=(1+\mathbf V)/2-\Psi-\mathbf E$ with $\mathbf{V}$ being the swapping operator of two quDits.
We note that
$\Psi,\mathbf E, \mathbf F$ and $\mathbf P$ are mutually orthogonal projections and $\mathbf F$ is a subspace the anti-symmetric subspace of two quDit with projector given by $(1-\mathbf V)/2$. Therefore $\mathbf K^2$ has a non-degenerated eigenvalue $4$ with eigenstate $|\Psi\rangle$, a degenerated eigenvalue $1$ with eigenspace $\mathbf E+\mathbf F$, a degenerated eigenvalue $4/d^2$ with eigenspace $\mathbf P$, and a degenerated eigenvalue $0$ with eigenspace $(1-\mathbf V)/2-\mathbf F$. Furthermore it can be checked that $\mathbf K|f_\tau\rangle=-|f_\tau\rangle$ while $\mathbf K|e_\tau\rangle=|e_\tau\rangle$ and $\mathbf K|\Psi\rangle=2|\Psi\rangle$. As a result the spectrum of $\mathbf K$ is $\{2,\pm 1,0,\pm 2/d\}$ and the eigenvalue $2$ is non-degenerated with eigenstate $|\Psi\rangle$ and  eigenvalue $-1$ is degenerated with $D$ eigenstates $|f_\tau\rangle$ in the case of $d\ge 3$.
Thus $(2-\mathbf K)(1+\mathbf K)\ge0$ so that $\delta_{\Lambda^\prime}=0$ if and only if $|\Lambda\rangle=a|\Psi\rangle+\sum_{\tau}h_\tau | f_{\tau}\rangle$ for some $a,h_\tau$, i.e., the channel $\Lambda$ is a linear combination of the identity channel and a Hamiltonian channel with $H=\sum_{\tau}h_\tau\lambda_\tau/d$. In the case of $d\ge 3$ the channel $\Lambda$ preserves the commutativity if and only if $H=0$, i.e., the channel $\Lambda$ is exactly the cloning channel $\mathcal C$.

{\it Corollary } In the case of $d\ge 3$ a mixing $p\mathcal I+\bar p \Lambda$ ($1>p>0$) of an arbitrary CoP channel $\Lambda$ with the identity channel $\mathcal I$ preserves the commutativity if and only if the channel $\Lambda$ is a cloning channel $\mathcal C(\varrho) \propto I\tr\varrho+c\varrho$.

The example of a unital non-CoP channel in the case of $d\ge 3$ given in \cite{2cp} is a mixing of the identity channel with a semi-classical channel, which is definitely not a cloning channel. Another example of non-CoP channel is the mixing of the identical channel with a unitary channel in the case of $d\ge 3$ while it preserves the commutativity in the case of $d=2$.

Finally, let us consider various quantitative measures for quantum correlations under the CoP channels. First of all since the CoP channels preserve the set of quantum classical states, all distance-based measures are non-increasing under CoP channels. In fact any such kind of measure is defined by the minimal distance to the set of classical states with some suitable distances such as the distance, geometric measure, or the relative entropy. There are two different definitions of quantum discord with one over orthogonal projective measurements and one over all possible measurements. It turns out that the quantum discord over von orthogonal projections does not have this desirable property. Let us consider the following 2-qubit state $\varrho=\sum_{a,b=0}^3R_{ab}\sigma_a\otimes\sigma_b$ with
\begin{equation}
R=\frac14\left(\begin{array}{cccc}1&1/4&-1/2&1/4\cr
1/4&2/5&0&0\cr
-1/6&0&1/5&0
\cr-1/20&0&0&-1/5
\end{array}\right)
\end{equation}
Its $A$-discord over orthogonal projections can easily be found numerically $0.0314231$ while for the state $\varrho_u=(\varrho+u_A\varrho u^\dagger_A)/2$ where the single qubit unitary transformation $u_A=\sin\frac{2\pi}5/\sqrt2+i\sigma_3\sin\frac{2\pi}5/\sqrt2-i\sigma_2\cos\frac{2\pi}5$ acts on subsystem $A$ only, the $A$-discord reads $0.0325923$ which is slightly larger. As to the case of quantum discord defined via minimization over all possible measurements no examples of increasing under local CoP channels is found so far. We conjecture that the quantum discord over general measurements, which can be different from that over von Neumann measurements even in the qubits cases \cite{chen}, is non-increasing under local CoP channels.

Note added. On finishing our manuscript Hu {\it etal.} obtained Theorem 1 \cite{hu}. The CoP channel $\Lambda(\varrho)\propto I\tr\varrho+c\varrho^T$ (with $|c|\le1$) provides a counterexample to their conjecture.

{\it Acknowledgement---}This work is supported by National
Research Foundation and Ministry of Education, Singapore (Grant No.
WBS: R-710-000-008-271) and NSF of China (Grant No. 11075227).


\begin{thebibliography}{99}
\bibitem{discord1} H. Ollivier and W. H. Zurek, Phys. Rev. Lett. \textbf{88}, 017901 (2001).
\bibitem{discord2} L. Henderson and V. Vedral, J. Phys. A \textbf{34}, 6899 (2001).
%\bibitem{others}A. R. Usha Devi and A. K. Rajagopal, Phys. Rev. Lett. \textbf{100}, 140502 (2008); S. Luo, Phys. Rev. A \textbf{77}, 022301 (2008); D. Girolami, M. Paternostro, and G. Adesso, arXiv:1008.4136; K. Modi \textit{et al.}, Phys. Rev. Lett. \textbf{104}, 080501 (2010).
\bibitem{dqc1} E. Knill and R. Laflamme, Phys. Rev. Lett. \textbf{81}, 5672 (1998); B.P. Lanyon, M. Barbieri, M.P. Almeida, and A.G. White, Phys. Rev. Lett. {\bf 101}, 200501 (2008).
\bibitem{datta} A. Datta, A. Shaji, and C. M. Caves, Phys. Rev. Lett. \textbf{100}, 050502 (2008).
\bibitem{int1}D. Cavalcanti et al. Phys. Rev. A 83, 032324 (2011); V.
Madhok and A. Datta, Phys. Rev. A 83, 032323 (2011).
\bibitem{2cp} A. Streltsov, H. Kampermann, and Dagmar Bru\ss, Phys. Rev. Lett. {\bf 107}, 170502 (2011).
\bibitem{oc}B. Daki\'{c}, V. Vedral, and \v{C}. Brukner, Phys. Rev. Lett. {\bf 105}, 190502 (2010).
\bibitem{pt}T. Werlang, C.Trippe, G.A.P. Ribeiro,and G. Rigolin, Phys. Rev. Lett. \textbf{105}, 095702 (2010).
\bibitem{CP} A. Shabani and D. A. Lidar, Phys. Rev. Lett. \textbf{102}, 100402 (2009); A. Shabani and D. A. Lidar, Phys. Rev. A \textbf{80}, 012309 (2009);
C. A. Rodr\'{\i}guez-Rosario \textit{et al.}, J. Phys. A \textbf{41}, 205301 (2008).
\bibitem{broadcast} M. Piani, P. Horodecki, and R. Horodecki, Phys. Rev. Lett. \textbf{100}, 090502 (2008);
S. Luo and W. Sun, Phys. Rev. A \textbf{82}, 012338 (2010).
\bibitem{cj}
Chengjie Zhang, Sixia Yu, Qing Chen, and C.H. Oh, Phys. Rev. A {\bf 84}, 032122 (2011).
\bibitem{yu1} Sixia Yu, Chengjie Zhang, Qing Chen, and C.H.Oh, arXiv: 1102.4710.
\bibitem{cs}B. Collins and P. Sniady, %"Integration with respect to the Haar Measure on the Unitary, Orthogonal and Symplectic Group."
Commun. Math. Phys. {\bf 264}, 773 (2006).% 795math-ph/0402073 (2004)
\bibitem{chen}Q. Chen, C. Zhang, S. Yu, X.X. Yi, and C. H. Oh, Phys. Rev. A {\bf 84}, 042313 (2011).
\bibitem{hu} X. Hu, H. Fan, D.L. Zhou, and W.-M. Liu, arXiv: 1112.3141.
\end{thebibliography}
\end{document}